\documentclass[aps,pre,onecolumn,eps,graphicx]{revtex4}
\def\bacol{\setlength{\arraycolsep}{0pt}}
\def\bec{\begin{center}}
\def\enc{\end{center}}
\def\be{\[}
\def\ben{\begin{equation}}
\def\ba{\begin{array}}
\def\bea{\begin{eqnarray}}
\def\ee{\]}
\def\een{\end{equation}}
\def\eea{\end{eqnarray}}
\def\ea{\end{array}}
\def\btab{\begin{table}}
\def\btabu{\begin{tabular}}
\def\etab{\end{table}}
\def\etabu{\end{tabular}}
\def\bit{\begin{itemize}}
\def\eit{\end{itemize}}
\def\bef{\begin{figure}[htb]}
\def\befh{\begin{figure}[!h!]}
\def\enf{\end{figure}}

\def\la{\langle}
\def\ra{\rangle}

\def\b1{{\bf 1}}

\def\cos{\hbox{cos}\:}

\def\cosh{\hbox{cosh}}

\def\nn{\nonumber}
\def\bb{\left(}
\def\eb{\right)}

\def\phip{{\phi^\prime}}
\def\alphap{{\alpha^\prime}}
\def\hpr{{h^\prime}}

\usepackage{epsfig,latexsym}

\newcommand \bew {\begin{widetext}}
\newcommand \enw {\end{widetext}}

\begin{document}

\title{\bf\noindent Weak non-linear surface charging effects in electrolytic films}

\author{D.S. Dean$^{(1)}$ and R.R. Horgan$^{(2)}$}

\affiliation{(1) IRSAMC, Laboratoire de Physique Quantique, Universit\'e Paul Sabatier, 118 route de Narbonne, 31062 Toulouse Cedex 04, France\\
(2) DAMTP, CMS, University of Cambridge, Cambridge, CB3 0WA, UK \\
E-Mail:dean@irsamc.ups-tlse.fr, rrh@damtp.cam.ac.uk}

\date{7 October 2002}
\begin{abstract}
A simple model of soap films with nonionic surfactants stabilized by 
added electrolyte is studied. The model exhibits charge regularization due
to the incorporation of a physical mechanism responsible for the formation
of a surface charge. We use a Gaussian field theory in the film but the 
full non-linear surface terms which are then treated at a one-loop level 
by calculating the  mean-field Poisson-Boltzmann solution and then the 
fluctuations about this solution. We carefully analyze the renormalization of the 
theory and apply it to a triple layer model for a thin film with Stern layer of 
thickness $h$. For this model we give expressions for the surface charge $\sigma(L)$ 
and the disjoining pressure $P_d(L)$ and show their dependence on the parameters. The influence
of image charges naturally arise in the formalism and we show that predictions depend 
strongly on $h$ because of their effects. In particular, we show that the surface charge vanishes as
the film thickness $L \to 0$. The fluctuation terms about this class of 
theories exhibit a Casimir-like  attraction across the film and although this attraction 
is well known to be negligible compared with the mean-field component for thick films 
in the presence of electrolyte, in the model studied here these fluctuations 
also affect the surface charge regulation leading to a fluctuation component
in the disjoining pressure which has the same behavior as the mean-field 
component even for large film thickness. 

\end{abstract}  
\maketitle
\vspace{.2cm}
\pagenumbering{arabic} 
\pagestyle{plain}

\section{\label{intro}Introduction}
Many situations in colloid, polymer and inter-facial science involve 
charged objects interacting in electrolyte solutions. In the case of
two interacting membranes or  soap films one encounters the  
electric double-layer. The mean-field theory of such experimental 
configurations is the Poisson-Boltzmann theory \cite{rusasc,is} for which
a surface boundary condition must be determined. 
For example, the surface charge 
or surface potential can be given, or a relation between the surface charge 
and surface potential may be specified. Such a relationship arises in a charge 
regularized model where the surface charging mechanism is derived from a microscopic 
description of the chemistry and geometry at the interface. Charge regularized 
models have the property that, owing to the thermodynamic nature of the charging 
mechanism, the surface charge changes as the distance between the two surfaces varies. 
The variation of the surface charge with the inter-surface distance will also
change the effective interaction between the surfaces and consequently also the
disjoining pressure. Within the film, the mean-field Poisson-Boltzmann theory is 
only sensitive to the electrostatic properties of the electrolyte; the chemistry and
effective sizes of the ions in the system only  enter into the description of the 
surface charging process. For example experiments 
\cite{sebe} shows that the disjoining pressure
increases with increasing hydration radius of the counter-ions in ionic soap films. 
This effect can be explained by the fact that as the counter ion radius increases  
the capacity for it to approach the surface and screen the surface charge is reduced.   

To take into account bulk surface tensions of electrolytes one must
resort to a microscopic description of the physics at the  interfaces. 
Depending on the nature of interface one may have specific adsorption due to 
chemical effects which can be taken into account via the law of mass action 
\cite{nipa1,cach,stjo,bebo,maar} or at a more statistical
mechanical level by introducing an external potential at the surface which
models the chemical liaison involved \cite{spbe2,dese,tapi}. In addition, 
there are other forces which come into play which are not present in the 
standard Poisson-Boltzmann approach. These are effects due to 
fluctuations in the electromagnetic field which can be identified 
with van der Waals forces \cite{lif,van,mani}. In bulk, the non-zero frequency 
van der Waals forces have little effect on the static ionic distributions since the 
relevant frequencies are too large and their contributions can be decoupled from 
those of zero frequency. However, it was pointed out in \cite{niya} 
that near interfaces the non-zero frequency van der Waals or dispersion forces can be 
important and depend strongly on the polarizability of the ions involved and
hence are ion specific. The zero frequency van der Waals forces, which correspond to 
thermal fluctuations in the electrostatic field, do strongly influence ionic 
distributions and so do modify the surface charge. 

When there are spatial variations in the dielectric constant image charges arise 
\cite{lali}. In the field theoretic approach adopted in this paper image charges 
and their effects are naturally and systematically included by taking into account 
the fluctuations of the electrostatic field.

In section \ref{model} we describe the field theory model we use to study the effects
of non-linear terms on the surface charging mechanism whilst retaining the free field 
theory description for the bulk electrostatic fields. This corresponds to using 
linear Debye theory in the bulk with fugacity $\mu$ but with the fully interacting 
description of the sources for the charging mechanism of the surfaces. The model is
applied to a triple layer thin film as an idealized model of the surfaces where there 
is a Stern layer of thickness $h$ from which all ions are excluded. In section
\ref{mft} we discuss the mean-field solution to the field theory for the thin film 
for which the non-linear interactions at the surface determine the source terms in the
mean-field equation. In section \ref{fluct} we give a detailed description of the effect
of field fluctuations about the mean-field solution using the Schr\"{o}dinger kernel
approach developed in an earlier paper \cite{deho}. We give predictions for the surface charge
$\sigma(L)$ and the disjoining pressure $P_d(L)$ as a function of film thickness $L$ and
show under reasonable assumptions that $\sigma(L) \to 0$ as $L \to 0$ and that the 
mean-field prediction for the large $L$ tail of $P_d(L)$ is modified by one-loop corrections,
both decaying like $\exp(-mL)$ for large $L$, where $m$ is the Debye mass
. In section \ref{discussion} we present a number of 
example graphs and discuss their salient features. In particular, it is clear that the
effects of image charges, which arise naturally and systematically in our formalism, are
very strong and that qualitative predictions depend sensitively on the value of $h$, the
thickness of the Stern layer. This demonstrates that a realistic model for the structure of
the surface and the Stern layer in particular is necessary for a quantitative study. We also
present some conclusions in this section.

\section{\label{model}Model}
We consider an idealized model of a thin film made with nonionic soap
adapted to the experimental set up used to measure the disjoining pressure as 
a function of the film thickness \cite{my,exkokh,dese,sebe,caetal}. 
The film shown Fig. (\ref{filmfig}) consists of two parallel surfaces of  area $A$ 
the interior of which is filled with a monovalent electrolyte solution
such as $NaCl$ in water with bulk dielectric constant $\epsilon$. The
exterior of the film is a dielectric medium of dielectric constant 
$\epsilon_0$, for example, air. The perpendicular distance between the 
two surfaces is denoted by $L+2h$. The region of thickness $h$ is the 
Stern layer from which the largest ions in the system are excluded
and $h$ can be taken to be the radius of the largest ion type in the system, which
is here chosen to be the anion (in most physical systems it is 
the cation with the largest radius due to hydration). If the radius of the 
cation in solution is $h'$ and $h'< h$ then the cation is excluded from a region 
of width $h'$ from the surface but can be present in the region $[-(h-h'),0]$ 
where there are no anions, thus leading to an effective surface charge in that region. 
Strictly speaking, just outside the film is the surfactant layer which in 
general will have a different dielectric constant to that of the  
exterior and the aqueous interior. Here for simplicity the presence of the 
nonionic surfactant is neglected. A version of this model with a single surface 
plus bulk is commonly used to model the surface properties of electrolyte solutions 
\cite{onsa,niya,lefl,kara}. In experiments the thickness $L$ of the
film may be varied by applying an external pressure in the cell containing the
film and its bulk. 

The grand partition function for this model system may be expressed as a 
functional integral 
 
\begin{equation}
\Xi = \int d[\phi] \exp\left(S[\phi]\right)
\end{equation}
where 
\begin{eqnarray}
S[\phi] = -{1\over 2} \int_{(T+L+2h)\times A} \beta \epsilon({\bf x}) (\nabla \phi)^2
d{\bf x} 
&+& 2\mu \int_{L\times A} \cos\left( \beta e \phi\right)d{\bf x}
\nonumber \\
&+& \mu_+^*  \int_{L\times A} (\delta(z) + \delta(L-z))\exp\left(i\beta e \phi\right)
d{\bf x}\;,
\label{actf}
\end{eqnarray}

where $e$ is the electron charge, $A$ is the area of the film and $\beta = 1/k_B T$. 
The fugacities of the anions and cations are taken to be equal and denoted by $\mu$
as are their densities denoted by $\rho$. The total length in the $z$ direction 
perpendicular to the film surfaces is $T+L+2h$ where $T$ denotes the total length 
external to the film. In the  region $z \in [-h,L+h]$ the dielectric constant, 
which is a function only of $z$, is given by $\epsilon(z) = \epsilon$ and outside 
the film $\epsilon(z) = \epsilon_0$.

The above field theoretical formulation can be obtained directly from Quantum 
Electrodynamics \cite{deho} by retaining just the electrostatic potential field in
the QED Lagrangian and coupling it to the distribution of ion charges. Alternatively, 
it can be obtained by standard field theory techniques based on the 
Hubbard-Stratonovich transformation for a monovalent Coulomb gas.  
The formulation takes into account both the Coulomb interactions between 
the ions and the zero frequency van der Waals forces due to fluctuations in
the electrostatic potential \cite{deho}. These zero frequency van der Waals 
forces are particularly relevant when there are variations in the dielectric 
constants of the system since they naturally and systematically include
the effects of image charges. This is especially the case for aqueous soap films 
in air where $\epsilon/\epsilon_0 \approx 80$. The density operators for the 
cations and anions can be shown to be
\begin{equation}
\rho_\pm({\bf x}) = \mu \exp\left(\pm i e \beta \phi({\bf x})\right)
\label{rho}
\end{equation}
The last term in Eq. (\ref{actf}) represents a highly 
localized affinity for the cations to be at the surfaces $z=0$ and $z=L$ 
of the film, and is responsible for the generation of a surface charge. 
There are various mechanisms leading to affinities for ionic species at 
interfaces, ranging from chemical affinity to steric and entropic 
effects. Here this term arises in the following approximation. Since there
can be cations in the region $[-(h-\hpr),0]$ and the corresponding region 
$[L,L+(h-\hpr)]$ there is, in addition to the first two terms of the action 
Eq. (\ref{actf}), a surface term
\begin{equation} 
\Sigma = \mu \int_{[-(h-\hpr),0]\times A} \exp\left( i\beta e \phi\right)d{\bf x}
+ \mu \int_{[L,L+(h-\hpr)]\times A} \exp\left( i\beta e \phi\right)d{\bf x}\;.
\end{equation}
This is the integral of the density operator given by Eq. (\ref{rho}) for the 
cations over the regions $[-(h-\hpr),0]$ and $[L,L+(h-\hpr)]$. These are the regions
in the Stern layers which may be occupied by the cations but from which the anions 
are excluded. If $h-\hpr \ll l_D$, where $l_D$ is the Debye length which 
characterizes the scale of the electrostatic interactions, then we can approximate 
$\Sigma$ by
\begin{equation}
\Sigma = \mu(h-h') \int_{L\times A} \delta(z)\exp\left( i\beta e \phi\right)d{\bf x}
+ \mu(h-h') \int_{L\times A}\delta(z-L)  
\exp\left( i\beta e \phi\right)d{\bf x}
\end{equation}
Comparing with the formula Eq. (\ref{actf}) we see that $\mu^*_+ = \mu(h-h')$.
In general  actions of the type Eq. (\ref{actf}) can be used to describe
various surface charging mechanisms with the proviso that the region where
the surface charge is localized has a width much smaller than the Debye length, $l_D$.
Whilst this description acts as a motivation for the model it is clear that in
general there will in general be specific adsorption for the different species 
near the air water interface. To model a specific adsorption
for the cations in the surface region $[-(h-\hpr),0]$ we
take a surface cation fugacity $\mu_s$ in this region 
which is greater than the bulk cation fugacity $\mu$. On shrinking the surface 
region to zero one would then have $\mu^*_+ = \mu_s(h-\hpr)$. There is some evidence 
for specific absorption of certain ionic species at seemingly chemically neutral interfaces; 
the famous and controversial Jones-Ray dip \cite{jreff} in the surface tension of 
weak electrolyte solutions (with interfaces with air) can  be explained by invoking a 
specific adsorption of anions at the interface \cite{dole}, although the basic surface exclusion 
model introduced in \cite{onsa} and used here cannot explain negative excess surface 
tensions. Recent experimental evidence points towards a specific adsorption of hydroxide
ions at air/water and oil/water interfaces \cite{kara}. In what follows we treat $\mu^*_+$
as an independent variable to account for the more complex charging mechanisms which can     
occur at the interface.

It is important to note that the fugacities $\mu$ and $\mu^*$ are determined by the values
of the physical bulk density $\rho$ and surface density for a single planar
surface $\rho^*_+$. We then have 
\be
\rho = \mu\la \cos \beta e\phi \ra~,~~~~\rho^*_+ = \mu^*_+\la \exp(i\beta e\phi)\ra~,
\ee
where the brackets stand for averaging over the bulk partition function. In mean-field 
theory this implies $\rho = \mu,~\rho^*_+ = \mu^*_+$ but this is
not true in general and the relationship must be calculated taking field interactions
into account.

In \cite{deho} the field theory with the action of Eq. (\ref{actf}) was
analyzed in the weak-coupling or Debye-H\"uckel limit which is a Gaussian
approximation where the action is expanded to second order in the field $\phi$. 
This amounts to the assumption that the mean-field densities of cations and anions 
throughout the film are small enough so that $8\pi\rho(z)\,l_B^3 < 1$, where $l_B$ is the
Bjerrum length $l_B = e^2\beta/4\pi\epsilon$. Another approach is to solve the 
full non-linear mean-field equations and then calculate the one loop correction 
which gives the effect of field fluctuations about this mean-field solution 
\cite{atmini2,po}. In \cite{atmini2} the resulting mean-field solution in the case of 
fixed surface charge was ingeniously expressed in terms of special functions
which allowed an analytic calculation of the one loop correction. In our case, however, 
the fact that the surface charge must be computed self consistently leads to 
additional complications and we take a different approach. We assume that the 
Gaussian approximation is valid inside the film but not at the surface where, because of
the increased charge density, the electrostatic interactions will be stronger and 
we retain the full non-linear surface operators. The resulting theory can be analyzed
as before \cite{deho} but with an effective $L-$dependent surface source for the field. 
The theory is accurate to the same order in perturbation theory as before but, in addition, 
now includes non-perturbative surface effects. The corresponding Debye-H\"uckel action 
with non-linear surface terms is 
 
\bacol{
\begin{eqnarray}
S^*&&[\phi] ~=~ -{1\over 2} \int_{(T+L+2h)\times A} \beta \epsilon(z) (\nabla \phi)^2
d{\bf x} - {1\over 2}\int_{L\times A} \beta \epsilon m^2 \phi^2 \ d{\bf x}
\nonumber \\ +~
&&\mu_+^*  \int_{L\times A} (\delta(z) + \delta(L-z))\exp\left(i\beta e \phi\right)
d{\bf x} + 2\mu A L \;,
\label{actdh}
\end{eqnarray}
}
where $m = \sqrt{8\pi\rho\,l_B} \equiv 1/l_D$ is the bulk Debye mass, $l_D$ is 
the Debye length and $l_B$ is the Bjerrum length defined above. The weak coupling limit 
corresponds to $m\,l_B < 1$, and the Gaussian approximation will be valid throughout the 
film so long as the local or effective mean-field Debye mass $m(z) = \sqrt{8\pi\rho(z)\,l_B}$ 
does not become so large that this weak coupling condition is violated. This condition 
is $8\pi\rho(z)l_B^3 < 1$ as stated earlier. We note that since $\rho(z) > \rho_{bulk}$ this
necessarily requires that $8\pi\rho_{bulk}\,l_B^3 < 1$.

Proceeding with the approximation scheme described above we have the expression
\begin{equation}
\Xi \approx  \int d[\phi] \exp\left(S^*[\phi]\right)
\end{equation}
for the grand partition function. The mean-field Poisson-Boltzmann equation is 
obtained from the saddle point of the action $S^*$:
\begin{equation}
{\delta \over \delta \phi({\bf x})}S^*\vert_{\phi_c} = 0 \label{mf}
\end{equation}
with $\mu^*_+$ replaced by $\rho^*_+$. This is the correct procedure since 
perturbative corrections due to the interactions of field fluctuations about this 
mean-field, or tree level, solution will then relate $\rho^*_+$ and $\mu^*_+$
self-consistently order by order. In principle, the same approach applies to the
bulk quantities $\rho$ and $\mu$ but since we are limiting the current analysis to
a Gaussian theory in the bulk, the distinction need not be made at this stage.

Taking into account the Gaussian fluctuations about the mean-field solution gives the
perturbative correction to one-loop for which we have
\begin{equation}
\Xi \approx \exp\left(S^*[\phi_c]\right)\int d[\phip]
\exp\left({1\over2}\int d{\bf x} d{\bf y} {\delta^2S^*\over 
\delta \phi({\bf x}) \delta \phi({\bf y})}\vert_{\phi_c} \phip({\bf x})\phip({\bf y
}) \right)\;,
\end{equation}
with, again, $\mu^*_+$ replaced by $\rho^*_+$. All other terms are treated as interactions
to be analyzed by perturbation theory.

The grand-potential per unit area of film $J$ can be separated into a mean-field 
contribution plus the zero frequency van der Waals contribution coming from the 
fluctuations. We write
\begin{equation}
J = J^{MF} + J^{vdW}
\end{equation}
where
\begin{equation}
J^{MF} = -{1\over A\beta} S[\phi_c]
\end{equation}
and
\begin{equation}
J^{vdW} =  {1\over 2 A\beta} {\rm Tr}\ln\left(-{\delta^2S^*\over 
\delta \phi({\bf x}) \delta\phi({\bf y})}\vert_{\phi_c}\right)
\end{equation}
\section{\label{mft}Mean-Field Theory}
The mean-field equation is obtained as  usual by  looking for an 
imaginary solution to Eq. (\ref{mf}) $\phi_c = i \psi$ where $\psi$ is 
real and corresponds to the mean-field electrostatic potential \cite{deho}.
The resulting equation for $\psi$ is
\begin{equation}
\beta\nabla\cdot \epsilon \nabla \psi - m^2 \beta \epsilon \psi
+ \rho_+^* \beta e  (\delta(z) + \delta(L-z))\exp\left(-\beta e \psi\right)~=~0\;,
\end{equation}
within the film and outside the film one has
\begin{equation} 
\beta\nabla\cdot \epsilon_0 \nabla \psi = 0\;. \label{mfout}
\end{equation}
This mean-field equation has the form of a standard linearized Poisson-Boltzmann 
equation but with non-linear boundary terms with $\mu^*_+$ is replaced by its 
mean-field approximation $\rho^*_+$ as explained in the previous section.  

The solution for $\psi$ is by symmetry  only dependent on
$z$ and symmetric about the mid-plane of the film at $z = L/2$ and so
we choose the solution $\psi(z) = C(L) \cosh\left(m(z - {L\over 2})\right)$
inside the film. Outside the film Eq. (\ref{mfout}) gives that
$d\psi/dz = 0$, which is the  condition of electro-neutrality of the mean-field
solution within the film. Integrating the mean-field equation 
between $z= 0^-$ and $z= 0^+$, and using the condition of electro-neutrality, one finds
\begin{equation}
\epsilon {d \psi\over dz}|_{z= 0^+}
= -e\rho_+^*\exp\left(-\beta e \psi(0)\right)
\end{equation}
Defining $D(L) =  C(L)\,\beta e\,\cosh(mL/2)$ gives the non-linear 
self consistent equation determining $D$ to be
\begin{equation}
D(L) = \alpha\coth({mL\over 2})\exp(-D(L)) \label{bcnl}
\end{equation}
where $\alpha = m\rho^*_+/2\rho$ is dimensionless. In the Stern layer model 
$\alpha = m(h-h')/2 = (h-h')/2 l_D$.  As stated in Section \ref{model} it is 
when $l_D \gg (h-h')$ that the formulation in terms of a surface charge is valid. 
In the study of this particular model one has that $\alpha \ll 1$, which also 
implies small surface charges compatible with the use of the quadratic 
approximation within the film. Then  Eq. (\ref{bcnl}) may be formally solved as a 
power series in $ \alpha\coth({mL\over 2})$ using standard series inversion
techniques from complex analysis. We find that 
\begin{equation}
D(L) = \alpha \coth\left({mL\over 2}\right)\sum_{n=0}^\infty (-1)^n
\left(\alpha \coth\left({mL\over 2}\right)\right)^n {(n+1)^{n-1}\over n!}
\label{dseries}
\end{equation}
We note that in general surface charge regulated models \cite{bebo},
even if there is an exact solution to the Poisson-Boltzmann equation 
(or its linearized form) in the bulk, one must determine the surface 
potential via a transcendental equation relating the surface charge to the 
surface potential which in general may be solved numerically by iteration
\cite{nipa1} or by linearizing the boundary equation \cite{cach}. 
Fortunately in the case studied here we have an explicit series solution to 
the boundary equation. The disjoining pressure of a film $P_d(L)$ is the 
difference between the film and bulk pressures. In the grand canonical ensemble

\begin{equation}
P_d(L) ~=~P(L) - P_{{\rm bulk}} ~=~ -{\partial J(L)\over \partial L} + 
\lim_{L\to \infty} J(L)/L 
\end{equation}
where $J$ is the film grand potential per unit area. We can decompose the disjoining 
pressure into a contribution coming from the mean-field solution and
a contribution coming from the field fluctuations which corresponds to the zero 
frequency van der Waals interaction
\begin{eqnarray}
P^{MF}_d(L) &=& -{\partial J^{MF}(L)\over \partial L} + \lim_{L\to \infty} J^{MF}(L)/L
\\
P^{vdW}_d(L) &=& -{\partial J^{vdW}(L)\over \partial L} + \lim_{L\to \infty} J^{vdW}(L)/L
\end{eqnarray}
After some straightforward but laborious algebra one finds that
\begin{equation}
P_d^{MF}(L)  = \rho k_B T {D^2(L)\over \cosh^2\left( {mL\over 2}\right)}
\label{eqpmf}
\end{equation} 
The mean-field value of the density within the film is given by
\begin{equation}
\rho_{MF}({\bf x}) = 2\rho\,\cosh\left(e\beta \psi({\bf x})\right)\;,
\end{equation} 
and in the linearized theory within the film this becomes
\begin{equation}
\rho_{MF}({\bf x}) = 2\rho + \rho e^2\beta^2 \psi({\bf x})^2\;. \label{potmp}
\end{equation}
At the mid-plane $z = L/2$ of the film one has
\begin{equation}
\rho_{MF}(z = {L\over2})) - \rho_{\rm bulk} = \rho\,{D^2(L) \over \cosh^2({mL\over 2})}\;.
\label{dpmp}
\end{equation} 
The mid-plane pressure formula \cite{rusasc,is} for Poisson-Boltzmann theories with
fixed surface charges or potentials, relates the disjoining pressure 
to the mid-plane mean-field density by  
\begin{equation}
P_d^{MF} = k_B T [ \rho_{MF}(z = {L\over2}) - \rho_{\rm bulk}]\;. \label{mpnf}
\end{equation}
In fact, in theories of the type considered here with potentials at or near the film 
surface, one can show quite generally that the mid-plane formula
holds generically as long as the external potential is zero in a finite interval
containing the mid-plane $z=L/2$. One can see directly, comparing
Eq. (\ref{eqpmf}) and Eq. (\ref{mpnf}) with the linearized approximation
Eq. (\ref{dpmp}), that the mid-plane formula is respected here.

It is useful for what follows to work with the running variable
$\alphap(L) =  D(L)\,\tanh\left({mL\over 2}\right)$. The mean-field surface
charge $\sigma_{MF}(L)$ may be written in terms of $\alphap(L)$ 
\begin{equation}
\sigma_{MF}(L) = {2\rho e \alphap(L) \over m} \label{sigmf}
\end{equation} 
From Eq. (\ref{dseries}) we have 
\begin{equation}
\alphap(L) = \alpha \sum_{n=0}^\infty (-1)^n
\left(\alpha \coth\left({mL\over 2}\right)\right)^n {(n+1)^{n-1}\over n!}
\label{alseries}
\end{equation}
As $L\to \infty$, $\;\alphap$ takes its bulk surface value 
$\alpha^* = \lim_{L\to \infty} \alphap(L)$. As mentioned previously
we are using a Gaussian field theory inside the film and so the mechanism 
generating the surface charge cannot be taken to be too strong, implying
that $\alpha \ll 1$ which gives $\alpha^* \approx \alpha$. As
$L$ decreases $\alphap(L)$ decreases. However, since the expansion in  Eq. (\ref{alseries}) 
is in $\alpha\coth(mL/2)$, the non-linear terms in this series must be taken into account 
when varying $L$. Indeed, for small $L$ one can show that $\alphap(L)\sim -mL\ln(L)/2$, 
which tends to zero as $L$ tends to zero. 

The mean-field disjoining pressure in terms of $\alphap(L)$ is
\begin{equation}
P_{MF} = \rho k_B T {\alphap^2(L)\over \sinh^2({mL\over 2})}\;,
\end{equation}
which, in terms of the mean-field surface charge reads
\begin{equation}
P_{MF} = {\sigma_{MF}^2(L)\over 2\epsilon\sinh^2({mL\over 2})}\;.\label{psig}
\end{equation}
For the linearized Poisson-Boltzmann equation with constant surface charge 
$\sigma_c$ one finds that \cite{greg,maar}
\begin{equation}
P_{MF} = {\sigma_c^2\over 2\epsilon\sinh^2({mL\over 2})}\;.
\end{equation} 
Hence, at the mean-field level, fitting the disjoining pressure at each
value of $L$ with an $L$ dependent $\sigma_c$ will reproduce 
the behavior of $\sigma_{MF}(L)$ in the current theory. This result
is also true for full non-linear mean-field theory and is not 
dependent on the quadratic field approximation in the bulk used here.
 
\section{\label{fluct}Fluctuation Effects}
Evaluating the fluctuations about our mean-field solution yields 
\begin{eqnarray}
{1\over2}\int d{\bf x} d{\bf y} {\delta^2S_{DH}\over 
\delta \phip({\bf x}) \delta\phip({\bf y})}\vert_{\phi_c} \phip({\bf x})\phip({\bf y
}) &=& -{1\over 2} \int_{(T+L)\times A} \beta \epsilon(z) (\nabla \phip)^2
d{\bf x} - {1\over 2}\int_{L\times A} \beta\epsilon m^2 \phip^2 \ d{\bf x}
\nonumber \\
&&- {1\over 2} \beta \epsilon m\,\alphap(L)\int_{L\times A} (\delta(z) + \delta(L-z))\phip^2
d{\bf x}\;, \label{fluc}
\end{eqnarray}
where 
\begin{equation}
\alphap(L) = {D(L)}\,\tanh\left({mL\over 2}\right) = 
{m\rho+^*\exp\left(-\beta e \psi(0)\right)\over 2 \rho}\;. 
\end{equation}
The main difference in Eq. (\ref{jvdw}) from the pure Gaussian theory of \cite{deho} 
is that the surface term $\alphap(L)$ is now a function of the film thickness, whereas 
in the pure Gaussian theory it is a constant. The fluctuation term Eq. (\ref{fluc}) may 
be evaluated using functional techniques \cite{mani,atmini1,atmini2,kago,netz2} as it 
is a functional determinant, or by path integral techniques \cite{deho}. Using the 
results of \cite{deho} one finds that the terms depending explicitly on $\alphap$
and $L$ that will contribute to the disjoining pressure and the surface charge are
\begin{eqnarray}
\beta J^{vdW}(L,\alphap) &=&  \beta J^{vdW}_{\rm bulk} L +
{m^2\over 2 \pi}\int dk \ k
\ln\left(B(km) k + \alphap(L) + \sqrt{k^2 + 1}\right) \nonumber \\
&+& {m^2\over 4 \pi}\int dk \ k \ln\left( 1 - 
\left( {B(km)k + \alphap(L)  -\sqrt{k^2 + 1}
\over   B(km)k + \alphap(L) + \sqrt{k^2 + 1}}\right)^2 
\exp(-2Lm \sqrt{k^2 + 1})\right)\;, \label{jvdw}
\end{eqnarray}
where 
\begin{equation}
B(p) ={1 + \Delta\exp(-2ph)\over 1 - \Delta\exp(-2ph)}\;,
\end{equation}
with  $\Delta= (\epsilon_0 - \epsilon)/((\epsilon_0 + \epsilon)$.
The  term  $\beta J^{vdW}_{\rm bulk}$ in Eq. (\ref{jvdw}) is  the van der Waals
contribution to the the bulk grand potential per unit volume and is given by
\begin{equation} 
\beta J^{vdW}_{\rm bulk} = {m^2 \over 4 \pi}\int dk k\left( \sqrt{k^2 + 1}
-k\right)\;.
\end{equation}

The mean-field contribution to the bulk grand-potential per unit volume
is simply $\beta J^{MF}_{bulk} = -2 \mu$ since the mean-field solution in
the bulk is just $\psi = 0$; there are no surfaces to set up
a mean-field potential. At this point we may not replace $\mu$ by $\rho$ because we
need to work to one-loop for consistency. The total bulk grand-potential per 
unit volume is thus given by 
\begin{eqnarray} 
 \beta J_{\rm bulk} &=& -2 \mu + {m^2\over 4 \pi}\int dk 
k\left( \sqrt{k^2 + 1}-k\right) \nonumber \\
&=& -2 \mu - {m^3\over 12 \pi} + {1\over 8 \pi} \Lambda m^2 + 
O({m^4\over \Lambda}) \\\nn
&=&-2\mu - g{2\rho \over 3}\bb 1- {3\Lambda \over 2m}\eb \label{jb}
\end{eqnarray}
where $\Lambda$ is a momentum space cut-off corresponding to 
a short distance cut-off $a \sim 1/\Lambda$ and $g = m^3/8\pi\rho = m\,l_B$. 
We see that indeed the expansion is in the dimensionless coupling $g$ as asserted 
in section \ref{model}. The bulk density of electrolyte is given by
\be
2\rho~=~-\mu{\partial \over \partial \mu} \beta J_{\rm bulk}~=~ 
2 \mu + {m^3 \over 8 \pi} - {1\over 8 \pi} \Lambda m^2\;,
\ee
which can be written as
\be
\mu~=~Z\,\rho~,~~Z~=~1 - {g \over 2}\bb 1 + {\Lambda \over m}\eb~.
\ee
From its definition in section \ref{model} we then also have, to this order in $g$,
that $\mu^*_+ = Z\,\rho^*_+$.

Substituting this result into Eq. (\ref{jb}) gives the well-known Debye
expression for the bulk pressure 
\begin{equation}
\beta P_{\rm bulk} = -  \beta J_{\rm bulk} = 2 \rho - {m^3 \over 24 \pi}\;.
\label{pbulk}
\end{equation}

In the field theoretic formulation used here the surface charge 
(on one surface) per unit area, $\sigma$, is given by
\begin{equation}
\sigma = e\mu^*_+\langle \exp(ie\beta\phi(z=0) \rangle\;, 
\end{equation}
which becomes, to the order of accuracy of the present treatment,
\begin{equation}
\sigma = e\mu^*_+\exp\left(-e\beta\psi(z=0)\right)\langle 
1 + i e \beta\phip(z=0) - {1\over 2} e^2 \beta^2\phip^2(z=0)\rangle\;,
\label{scr}
\end{equation}
where the average  $\langle \cdot \rangle$ in the above equation is
over the fluctuations $\phip$. As the one-loop action in $\phip$ is 
quadratic, the average of the term linear in $\phip$ in Eq. (\ref{scr}) is
zero and we may then write
\begin{equation}
\sigma = {2\mu e \alphap(L) \over m} - {e\alphap(L)\over 2} {\partial  \beta J^{vdW}
\over \partial \alphap(L)}\vert_{\alphap = 0} + O(\alphap^2)\;, 
\label{sc}
\end{equation}
where we have used that $\alpha = m\rho^*_+/2\rho = m\mu^*_+/2\mu$ 
and where we have kept only the leading order behavior of the surface 
charge in $\alphap$. The next order terms can be calculated and are finite 
though one needs to eliminate certain artificial divergences \cite{dehoprep}. 
The formula Eq. (\ref{sc}) gives a surface charge susceptibility with respect 
to the conjugate variable $\alphap$. We note that the first term in on the 
right hand-side of Eq. (\ref{sc}) is simply the mean-field contribution to the 
surface charge $\sigma_{MF}$. 

To $O(\alphap)$ we obtain
\begin{eqnarray}
\sigma(L) &=&  {2 e \alphap(L) \over m} \left( \mu - {m^3 \over 8 \pi}
\int {kdk\over D_+(k,m)}\right. \nonumber \\
&+&\left. {m^3 \over 16\pi} \int kdk\; {D_-(k,m)\over 
D_+(k,m)\;[D_+(k,m)^2 \exp(2Lm\sqrt{k^2+1}) + D_-(k,m)^2]} \right)
+O(\alphap^2)\;, 
\label{sig}
\end{eqnarray}
where $D_\pm(k,m) = k B(km) \pm \sqrt{k^2+1}\;$. There is, however, a divergence 
in the first integral term in Eq. (\ref{sig}). This term corresponds to the 
van der Waals contribution to the  charge of a bulk surface 
$\sigma_{\rm bulk} = \lim_{L\to \infty}\sigma$. This divergence can be regularized 
by choosing the fugacity $\mu$ to give the desired bulk electrolyte density 
$\rho$ correct at one loop order. We define
\begin{equation}
\Gamma =\int kdk\left({1\over k B(km) + \sqrt{k^2 + 1}} -{1\over 2k}\;,
\right)
\end{equation}
and then 
\begin{equation}
\sigma_{\rm bulk} = {2 e \alpha^* \over m} \left( \mu - {m^3 \over 8 \pi}
\Gamma - {m^2 \Lambda\over 16 \pi} \right)\;, 
\end{equation}
where $\Lambda$ is the ultra-violet cut-off introduced in the bulk calculation above. 
Note that for weak charging $\alpha^* \sim \alpha$. Hence in terms of the physical 
variable $\rho$ we obtain the divergence-free formula for $\sigma_{\rm bulk}$
\begin{equation}
\sigma_{\rm bulk} = {2 e \alpha^* \over m} \left( \rho- {m^3\over 16 \pi}- {m^3 \over 8 \pi}
\Gamma \right)\;. \label{sigb}
\end{equation}
We remark here that in the case $\Delta \neq 0$, if the Stern layer thickness 
$h$ is taken to zero then the term $\Gamma$ in Eq. (\ref{sigb}) diverges. 
It is clear however that one cannot have a surface charge exactly at the interface 
between two media of different dielectric constants due to the presence
of arbitrarily close image charges. The divergence in $h$ as $h \to 0$
in Eq. (\ref{sigb}) is thus a physical divergence and any model using a
surface charge must place this surface charge away from a discontinuity
in the dielectric constant. When $\Delta = 0$ no such divergence is present
and $B(km) = 1$ leading to  the simple formula 
\begin{equation}
\sigma_{\rm bulk}=  \left( 1 - {g \over 6}\right)\sigma_{MF}(\infty)\;.
\end{equation}
We notice that from Eq. (\ref{pbulk}) that this equation may be written 
as 
\begin{equation}
\sigma_{\rm bulk}=  {\beta P_{\rm bulk}e \alpha^* \over m}\;.
\end{equation}
From Eq. (\ref{sigb}) we see that effect of electrostatic interactions 
is to reduce the surface charge from the value it would have had without interactions.
This is because the excess anions left in the bulk pull the cationic surface charge 
into the bulk. The case where $\Delta > 0$ ({\em i.e.} $\epsilon_0 > \epsilon$) 
leads to $B(km) \ge 1$ for all $k$ and, examining the integrand in the formula 
defining $\Gamma$, we find that $\Gamma <0$ and hence that positive $\Delta$ increases
the surface charge above that of the case $\Delta = 0$. This is to be expected
physically as the image charges in this case attract the ions towards
the medium of higher dielectric constant \cite{lali}. In the case where 
$hm \ll 1$ we may evaluate the integral defining $\Gamma$ since the leading
divergence as $hm \to 0$ comes from the large $k$ integration. We find
\begin{equation}
\sigma_{\rm bulk} =  
\left( 1 - {g \over 6} + {g\Delta \over 4mh}\right)\sigma_{MF}(\infty)\;. \label{sigbd}  
\end{equation}
Again, we see that for  $\Delta > 0$ ($\Delta < 0$) the enhancement (reduction) of 
the surface charge which can be physically attributed to the presence of image charges.

Finally the  $L$ dependence of the surface charge at $O(\alphap)$  is given by
\ben
\sigma(L)~=~\sigma_{MF}(L)\bb 1 - {g\over 2} - g\Gamma   
 + {g \over 2}\int kdk\; {D_-(k,m) \over 
D_+(k,m)\;[D_+(k,m)^2 \exp(2Lm\sqrt{k^2+1}) + D_-(k,m)^2]}\: \eb\;, 
\label{sigl}
\een
where $D_\pm(k,m) = k B(km) \pm \sqrt{k^2+1}\;$.

Thus, because $\alphap(L) \to 0$ as $L\to 0$, we find that $\sigma_{MF}$, and hence the 
surface charge $\sigma$, vanishes as the film becomes thin. This is a physical result which 
is not picked up by a pure Gaussian theory \cite{deho}.

As previously stated the total disjoining pressure is composed of a mean-field 
contribution and a contribution coming from the fluctuations. We find the total 
of these two terms gives
\begin{equation}
P_d(L) = {\sigma^2_{MF}(L)\over 2 \epsilon \sinh^2({mL\over 2})}
\left( 1  + {2(\sigma(L) - \sigma_{MF}(L))\over \sigma_{MF}(L)
( 1 + \alphap(L)\coth({mL\over 2}))}\right)
- 4g\;\rho k_BT\int dk \ k \sqrt{k^2 +1)} {f^2(k) \over 1-f^2(k)}\;,
\label{pdtot}
\end{equation}
where
\begin{equation}
f(k) = {kB(km) + \alphap(L) - \sqrt{k^2 +1} \over kB(km) + \alphap(L) + 
\sqrt{k^2 +1}}\exp\left( - Lm \sqrt{k^2 +1} \right)\;.
\end{equation}
In the derivation of this formula the divergences which arise at intermediate stages
of the calculation can, as in the case of the surface charge, be shown to cancel
in the final result.

The first term in Eq. (\ref{pdtot}) is simply the mean-field contribution to 
$P_d$, the second (in the same bracket) is the contribution coming from the 
dependence of $J^{vdW}$ on $L$ via the $L$ dependence of $\alphap$. The last 
term is a form of screened zero frequency van der Waals contribution. In the absence 
of electrolyte we find that this last term gives a $1/L^3$ Casimir attraction
across the film. In the presence of electrolyte this interaction is screened
\cite{mani} and decays exponentially as $\exp(-2 m L)$ which is twice
as quickly as the mean-field contribution to the disjoining pressure which 
decays as $\exp(-mL)$ for large $L$. However in the theory presented here, the 
fluctuation effects give an additional term which decays also as $\exp(- m L)$
and hence the fluctuations modify the long distance behavior of $P_d$. We note 
theories with fixed surface charge the long-range component of the 
disjoining pressure is not altered at two-loop level and this effect
is specific to charge regularized models. 
The strength and sign of this long range modification of $P_d$
is controlled the  ratio  $  (\sigma(L) - \sigma_{MF}(L))/\sigma_{MF}(L)$ which
measures the relative deviation of the surface charge from its mean-field value. 
If the fluctuation effects enhance (reduce) the surface charge then the long range
value of $P_d$ is increased (decreased) from its mean-field value. 
The Casimir  term is, however, is always attractive.

It should be remarked at this stage that to this order, $O(g)$, there will be 
$L$-dependent contributions from the non-Gaussian interaction term in $\phip^4$. 
These contributions can be shown to vanish as $L \to \infty$ but their effect for
finite $L$ must be calculated. However, such a calculation requires the apparatus for
the general perturbation theory to be developed which we shall present in a forthcoming paper \cite{dehoprep}. 

\section{\label{discussion}Discussion and Conclusion}  

We have systematically developed a theory for a thin film with a full non-linear 
surface charging mechanism while retaining the free field theory description for
the bulk electrostatic fields. This corresponds to using linear Debye theory in 
the bulk with fugacity $\mu$ but with the fully interacting description of the 
sources for the charging mechanism of the surfaces. We have applied the theory to a 
model consisting of a triple layer system, shown in Fig. (\ref{filmfig}), in which 
there is adsorption of cations on to the surface modelled by a surface fugacity 
$\mu^*_+$ and encoded in the dimensionless surface absorption strength parameter 
$\alpha = m\rho^*_+/2\rho$ where $\rho$ and $\rho^*_+$ are the bulk density and cation 
density on the surface of a bulk region, respectively, and $m$ is the Debye mass. 
At the surface there is a Stern layer of thickness $h$ from which all ions are excluded. 
The film is of thickness $L$ and the dielectric constants are $\epsilon \sim 80\epsilon_0$ 
in the film and $\epsilon_0$ outside the film. This model for a real surface is too simple 
but it encodes the important feature that the thermodynamic properties of the film are very 
strongly dependent on the detailed nature of the surface and its properties. This is due to 
two features: the charging mechanism which allows the surface charge to remain in equilibrium 
with the interior charges and the effect of image charges due to the discontinuity in the 
dielectric constant at the surface. First we determine the mean-field solution 
$\phi_c(z)$ using the non-linear surface operators as the source and then we use the 
Schr\"{o}dinger kernel approach to calculate the partition function as an expansion in 
$\alpha$ and $g = ml_B$ where $l_B$ is the Bjerrum length. Much of the details of this
approach have been discussed in an earlier paper \cite{deho}. In this paper, we concentrate on
the effects of the non-linear surface charging mechanism which leads us to introduce an
effective, or running, surface charging parameter $\alphap(L)$ and we analyze the 
behavior of the surface charge $\sigma(L)$ and the disjoining pressure $P_d(L)$ on    
$m, \alpha$ and $h$. The formulas summarizing our findings are Eqs. (\ref{sigl}) and 
(\ref{pdtot}). Examples of the solutions to these equations are shown in 
Figs.
(\ref{p_disj_h}) to (\ref{sig_h_mf_1loop}) and we now briefly discuss the salient features of
these results.

In Figs. (\ref{alpha_a}) and (\ref{alpha_m}) we show $\alphap(L)$ as a function of $L$ for 
various values of $\alpha$ and $m$, respectively. $\alphap(L)$ controls the strength of
the surface charging mechanism and from Eq. (\ref{bcnl}), and what follows, it is clear that
$\alphap(L) \to 0$ as $L \to 0$ which in turn causes $\sigma_{MF}$ and $\sigma$ to vanish
also in this limit. 

In Figs. (\ref{p_disj_h}) to (\ref{p_disj_a}) we show the dependence of 
$P_d$ on $h,m,\alpha$, respectively. From all these figures we see that the characteristic
collapse transition is evident but that its strength is very sensitive to the parameters. 
In particular, from Fig. (\ref{p_disj_h}) we see that $P_d$ decreases as $h$ decreases, as
we should expect since the image charges at the surface are repelling the cations and so
reducing the surface charge. This effect can be seen directly in Fig.(\ref{sig_h}). The effect
on $P_d$ is due to the $\sigma$ dependent term in Eq. (\ref{pdtot}) which arises because of the
implicit $L$ dependence of the free energy through its dependence on $\alphap(L)$. Note that
$\sigma_{MF}(L)$ is independent of $h$ as in this formulation the 
effect of image charges first comes in at the one-loop level. 
The dependence of $P_d$ on $m$ and 
$\alpha$ shown
in Figs. (\ref{p_disj_m}) and (\ref{p_disj_a}) has the expected trends but again the height
of the peak in $P_d(L)$ is strongly dependent on the parameters which have been chosen 
to take values that can typically be achieved in experiment. In Fig. (\ref{p_disj_h_mf_1loop})
we show the  mean-field component and one-loop contributions 
for typical parameter values.  

In Figs. (\ref{sig_h}) to (\ref{sig_a}) the behavior of the surface charge $\sigma(L)$ is shown.
We see from Fig (\ref{sig_h}) that $\sigma(L)$ decreases very strongly with $L$ for small $L$
and that as $h$ decreases the effect strengthens and that for sufficiently small $h$ the value
of $\sigma(L)$ has a zero and becomes negative. Ultimately, $\sigma(L) \to 0$ as $L \to 0$
and so must have a minimum value before turning towards the origin. These effects are due to
the one-loop term in Eq. (\ref{sigl}) and can be seen clearly in 
Fig. (\ref{sig_h_mf_1loop}) where
the mean-field and one-loop contributions to $\sigma(L)$ are separately shown. In Figs.
(\ref{sig_m}) and (\ref{sig_a}) the trends shown are as expected but, as in the case of $P_d$ the
magnitude of $\sigma(L)$ is very sensitive to parameter values. The overall prediction is that
$\sigma(L)$ is strongly dependent on $L$ and vanishes as $L \to 0$.

In Fig. (\ref{p_disj_lin_nonlin}) we compare $P_d$ for the linearized theory from (\cite{deho}) 
and the non-linear theory of this paper with $h = 0.3~nm$. Although the peak in $P_d$ occurs in 
much the same place it is lower in the non-linear theory for this value of $h$. Since the peak height
is strongly dependent on $h$ we see that a quantitative prediction requires a realistic model
for the surface. See Fig. (\ref{sig_h}).

An important feature of these calculations is to note that the results are expressed as a 
series in both $\alpha$ or $\alphap(L)$ and $g = ml_B$ with partial resummations in some cases.
The major approximation is to use the free field theory within the bulk. The object was
to study the effects of the non-linear surface charging mechanism and we have shown that these
effects are indeed strong and it is clear that any approach which omits them or assumes a constant
surface charge will be incorrect. Some of the effects are strong and there are features, such as the 
minimum in $\sigma(L)$ and its change of sign, that must be studied further in the full non-linear
theory to see if they are not artifacts of the approximation. A consistent control
over spurious and artificial infinities must await the full perturbation 
theory. 
An example is
eluded to in Eq. (\ref{sc}) and what follows. We have indicated how to control such 
quantities here to the one-loop level and see that even here the analysis
is rather delicate. 
There are, in principle, $O(g)$ terms from interactions within the bulk which
will vanish as $L \to \infty$ but contribute finite $L$ effects to $P_d$ and $\sigma$ but these
terms are not expected to be large. We shall present an analysis of all these topics
in a forthcoming paper \cite{dehoprep} in which the full non-linear theory and its
perturbation expansion will be studied.

\befh
\bec
\epsfig{file=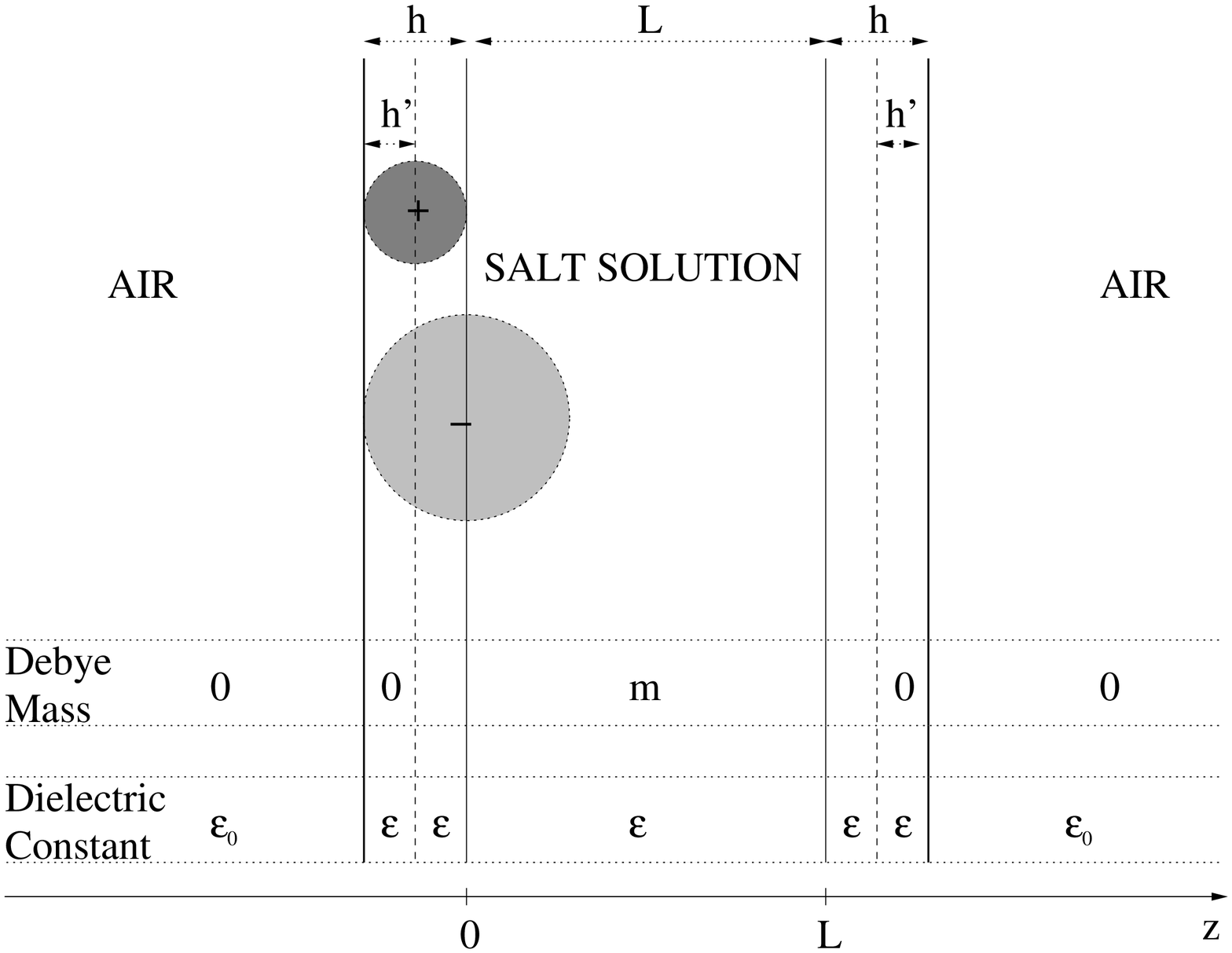,height=90mm}
\enc
\caption{\label{filmfig}\small 
Cross section through a model soap film. The distance of closest
approach of the ions to the surfaces (at $z=0$ and $z=L$) is the 
effective radius of the ionic species in solution. The dielectric 
constants and Debye masses as a function of the distance perpendicular
to the film surface ({\em i.e.})as a function of the position on the $z$ axis
are also shown.
} 
\enf

\befh
\bec
\epsfig{file=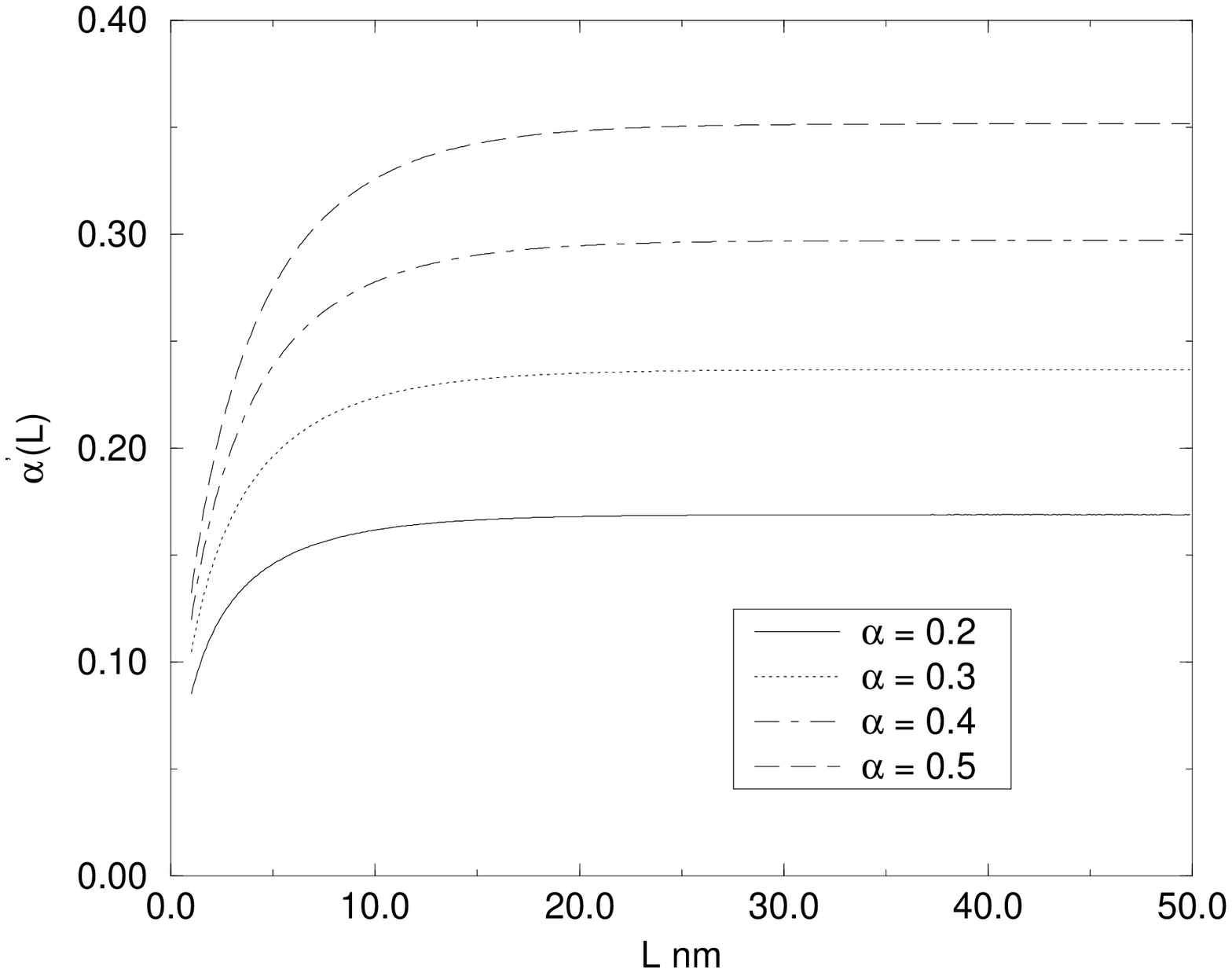,height=90mm}
\enc
\caption{\label{alpha_a}\small
$\alphap(L)$ for Debye mass $m=0.2~\mbox{nm}^{-1}$ and $h=0.3~\mbox{nm}$ for
values of $\alpha=0.2,0.3,0.4,0.5$ which controls the strength of the 
surface charging adsorption.
}
\enf

\befh
\bec
\epsfig{file=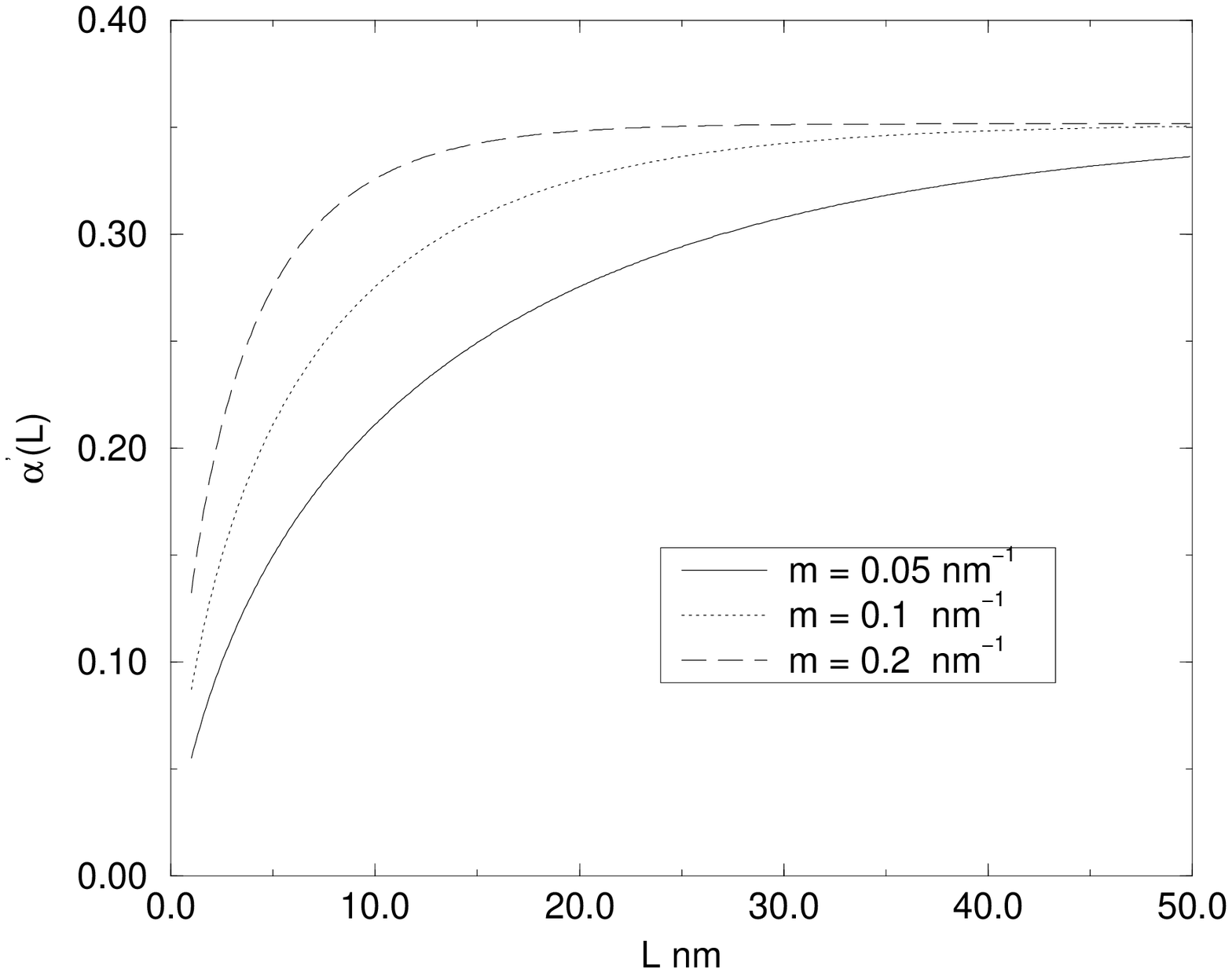,height=90mm}
\enc
\caption{\label{alpha_m}\small
$\alphap(L)$ for $\alpha=0.5$ and Stern layer thickness $h=0.3~\mbox{nm}$ for
values of Debye mass $m=0.05,0.1,0.2~\mbox{nm}^{-1}$. 
}
\enf

\befh
\bec
\epsfig{file=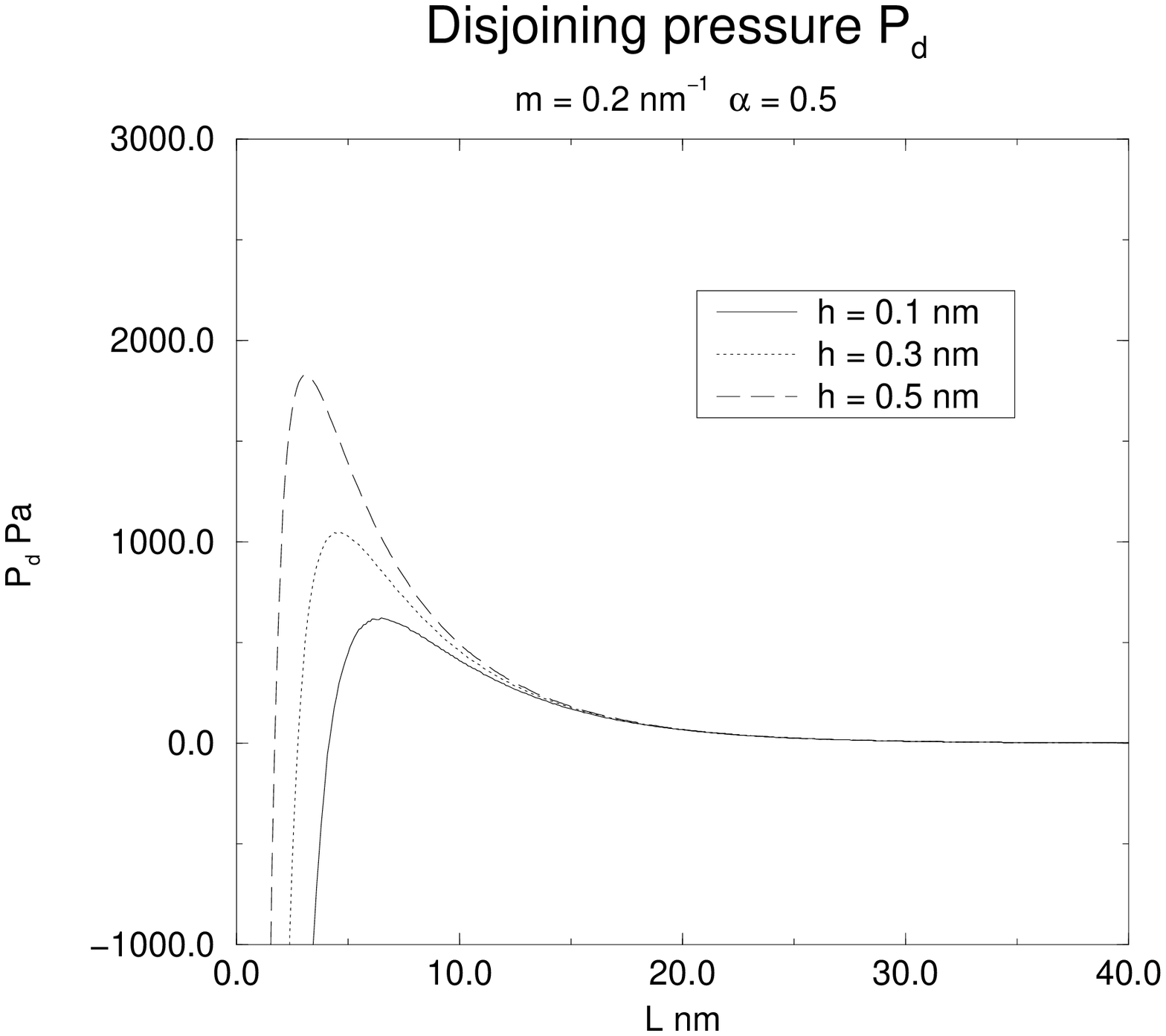,height=90mm}
\enc
\caption{\label{p_disj_h}\small
The disjoining pressure given in Eq. (\ref{pdtot}) for Debye mass 
$m=0.2 \mbox{nm}^{-1}$ and $\alpha = 0.5$ for different values of the thickness
of the Stern layer $h = 0.1, 0.3, 0.5~\mbox{nm}$. The sensitive dependence of $P_d$ on $h$
is evident as we should expect since the influence of the image charges increases
rapidly as $h$ decreases.
}
\enf

\befh
\bec
\epsfig{file=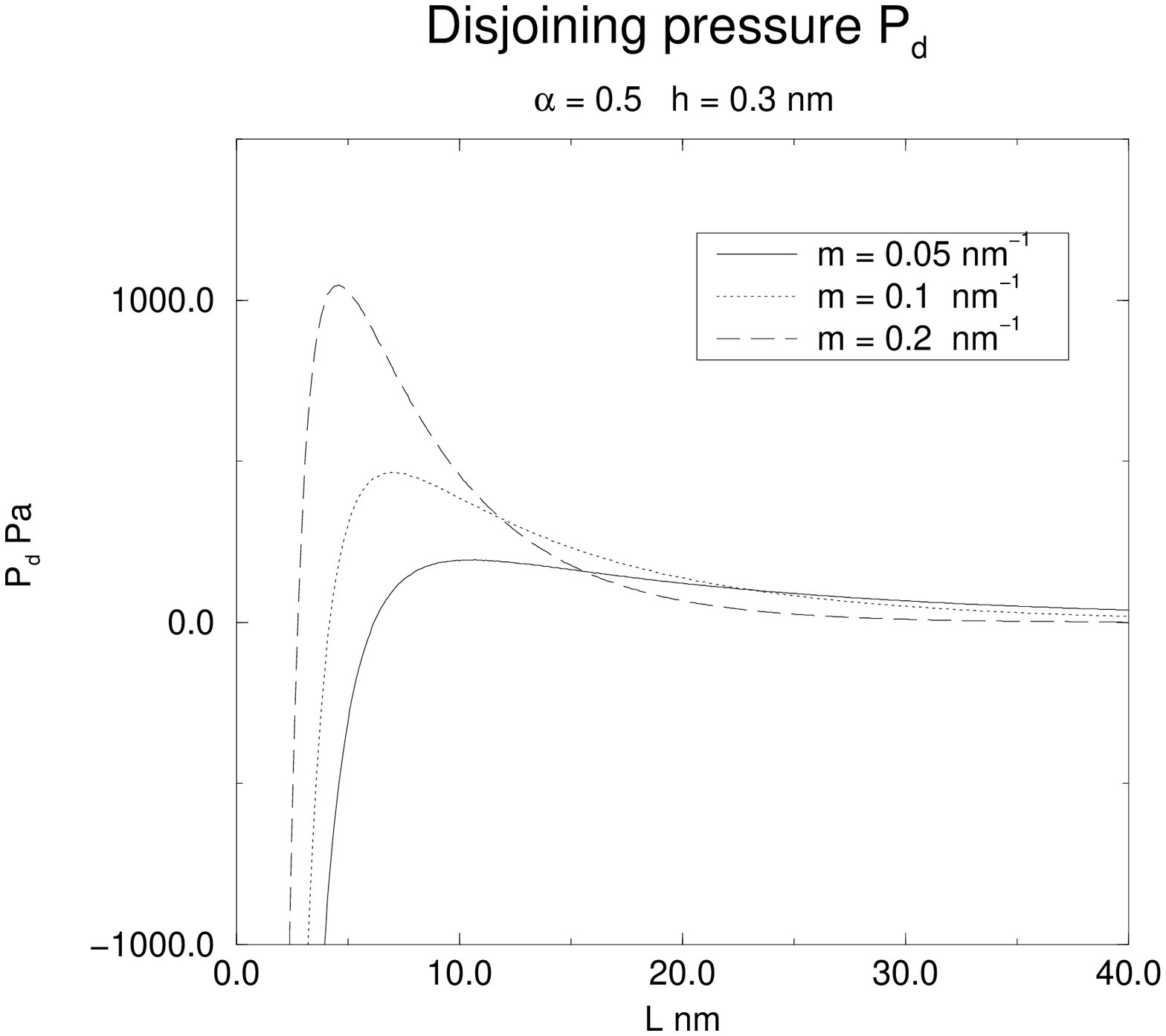,height=90mm}
\enc
\caption{\label{p_disj_m}\small
The disjoining pressure given in Eq. (\ref{pdtot}) for $\alpha=0.5$ and Stern layer
thickness  $h = 0.3~\mbox{nm}$ for values of the Debye mass $m=0.05,0.1,0.2~\mbox{nm}^{-1}$.  
}
\enf

\befh
\bec
\epsfig{file=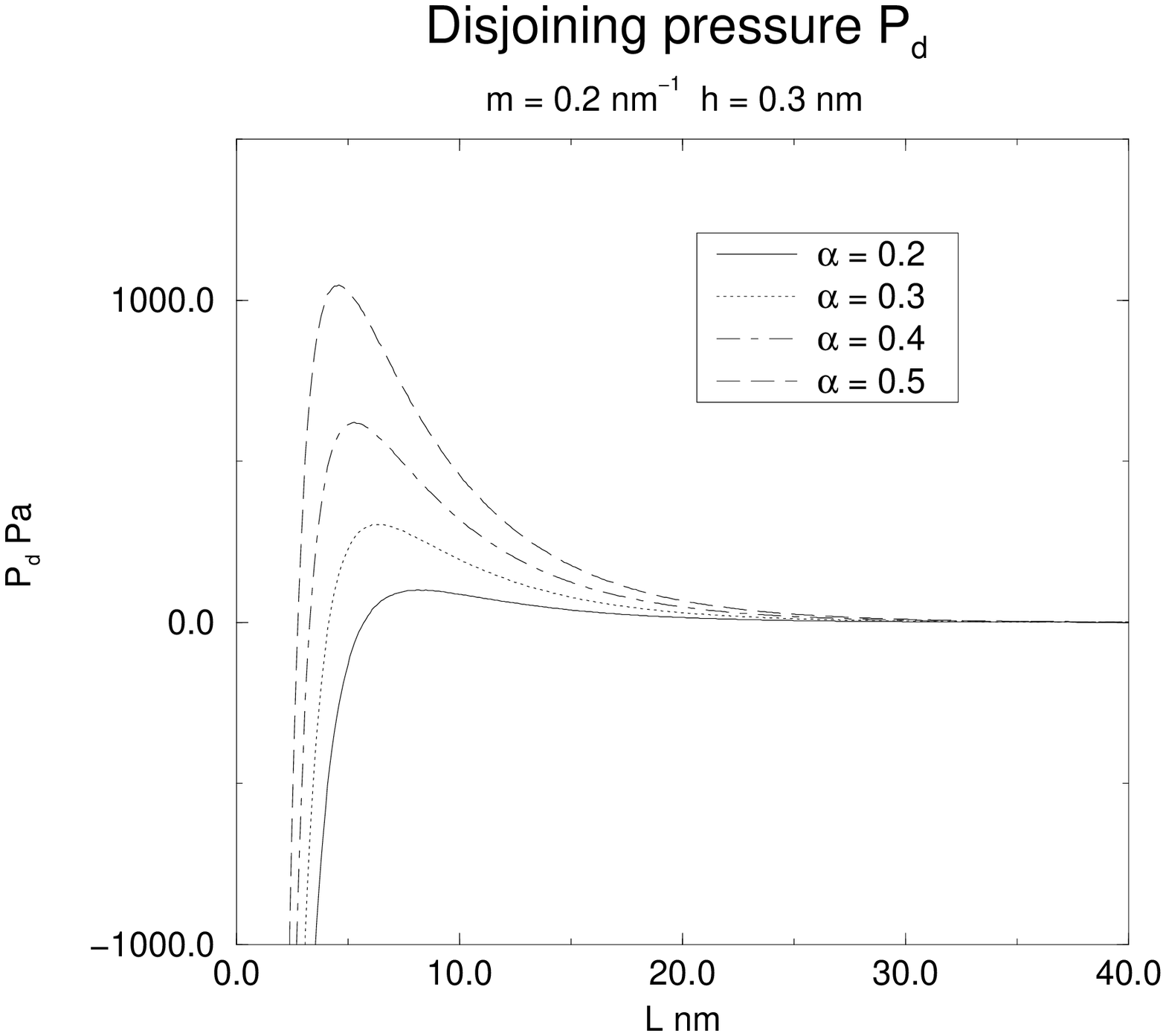,height=90mm}
\enc
\caption{\label{p_disj_a}\small
The disjoining pressure given in Eq. (\ref{pdtot}) for Debye mass $m = 0.2~\mbox{nm}^{-1}$
and Stern layer thickness  $h = 0.3~\mbox{nm}$ for values of $\alpha = 0.2,0.3,0.4,0.5$.
}
\enf

\befh
\bec
\epsfig{file=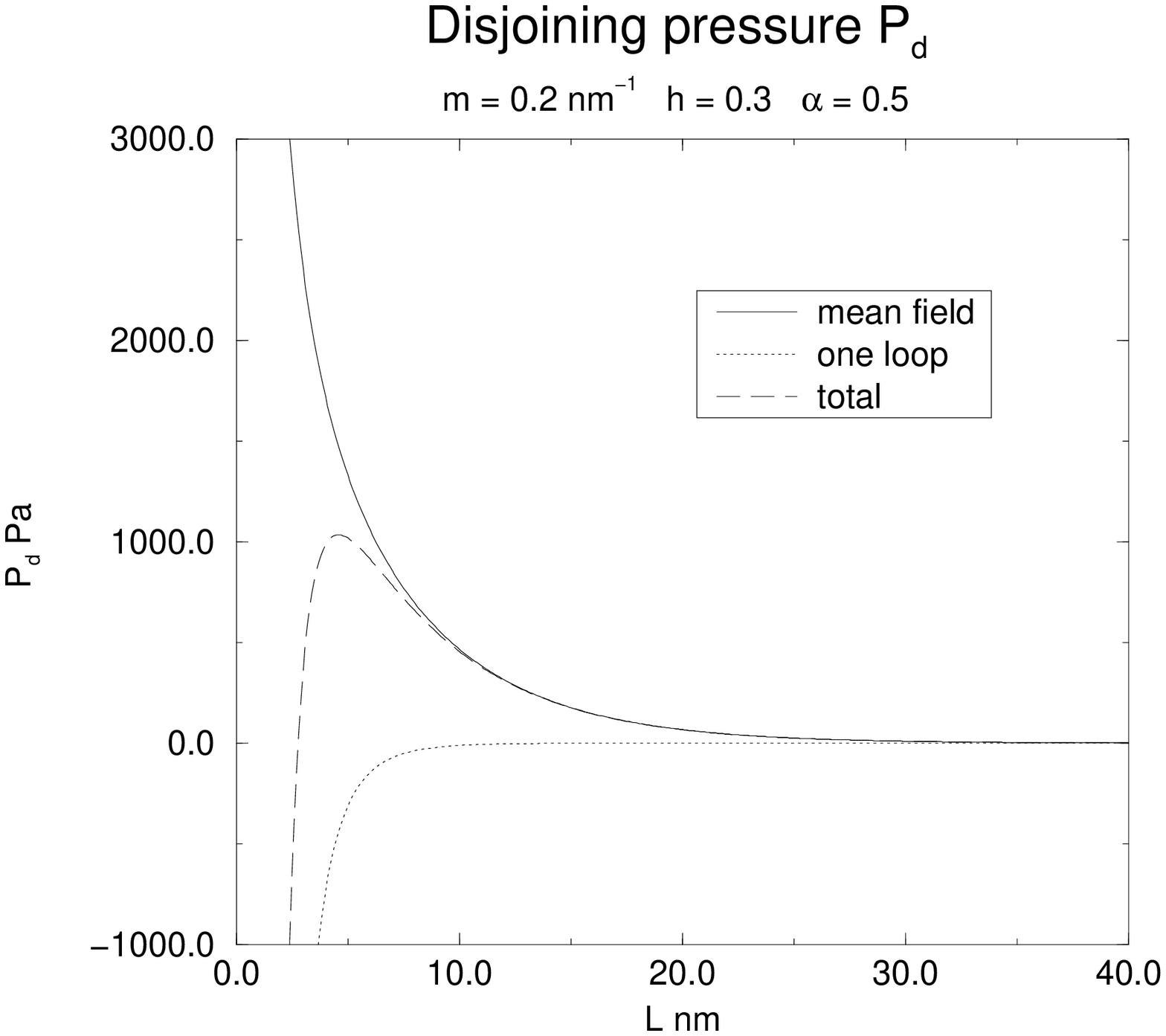,height=90mm}
\enc
\caption{\label{p_disj_h_mf_1loop}\small
The disjoining pressure given in Eq. (\ref{pdtot}) showing the mean-field contribution
and the one-loop ($O(g)$) contributions as a function of Stern layer thickness
$h =0.3$.
}
\enf

\befh
\bec
\epsfig{file=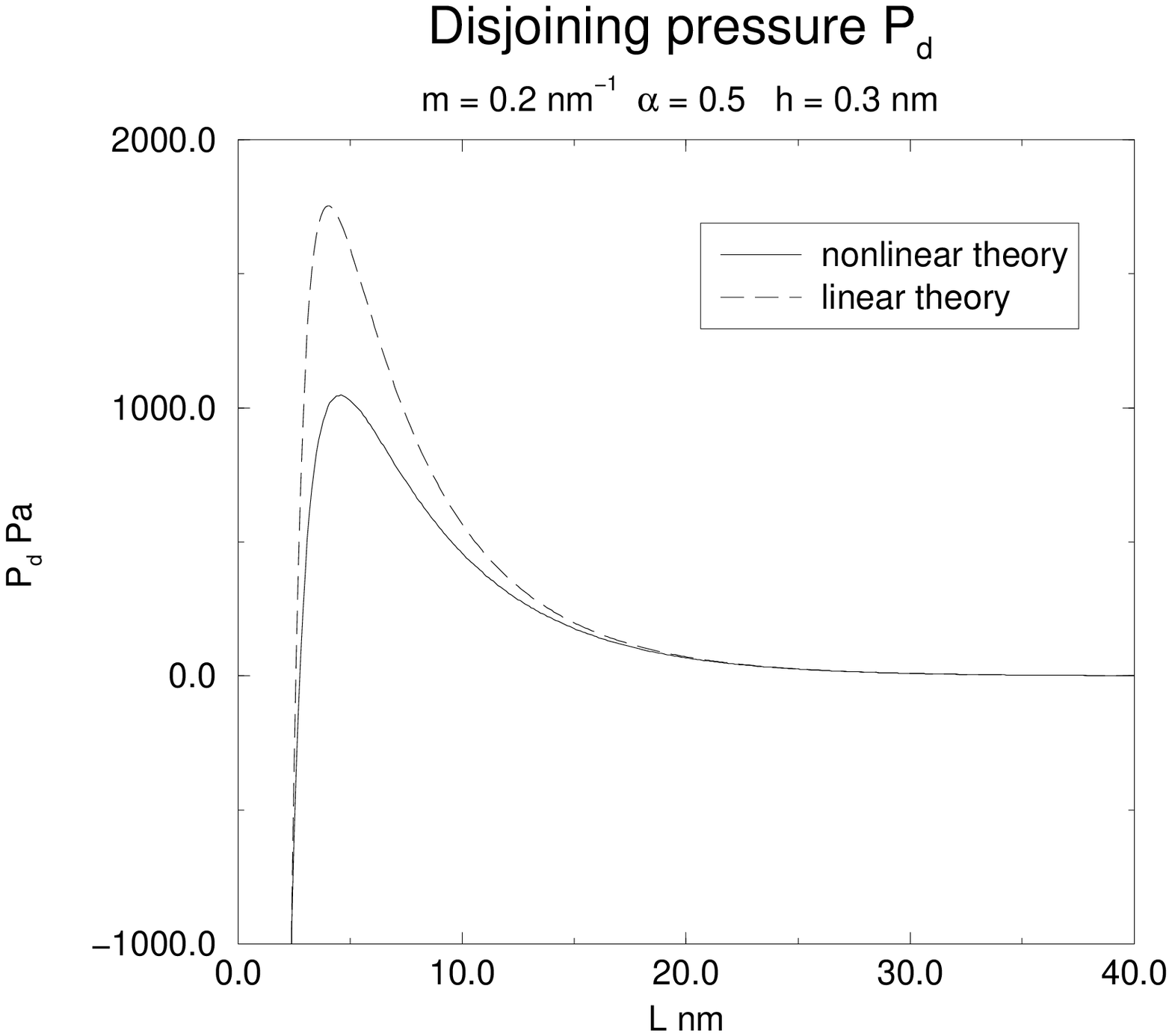,height=90mm}
\enc
\caption{\label{p_disj_lin_nonlin}\small
The disjoining pressure given in Eq. (\ref{pdtot}) for $m = 0.2~~\mbox{nm}^{-1}$
and $\alpha = 0.5$ for the linearized theory from (\cite{deho}) and the non-linear theory
of this paper for which $h = 0.3~nm$. Although the peak in $P_d$ occurs for much the same
value of $L$ it is lower in the non-linear theory for this value of $h$. Since the peak height
is strongly dependent on $h$ we see that a quantitative prediction requires a realistic model
for the surface. See Fig. (\ref{sig_h}). 
}
\enf

\befh
\bec
\epsfig{file=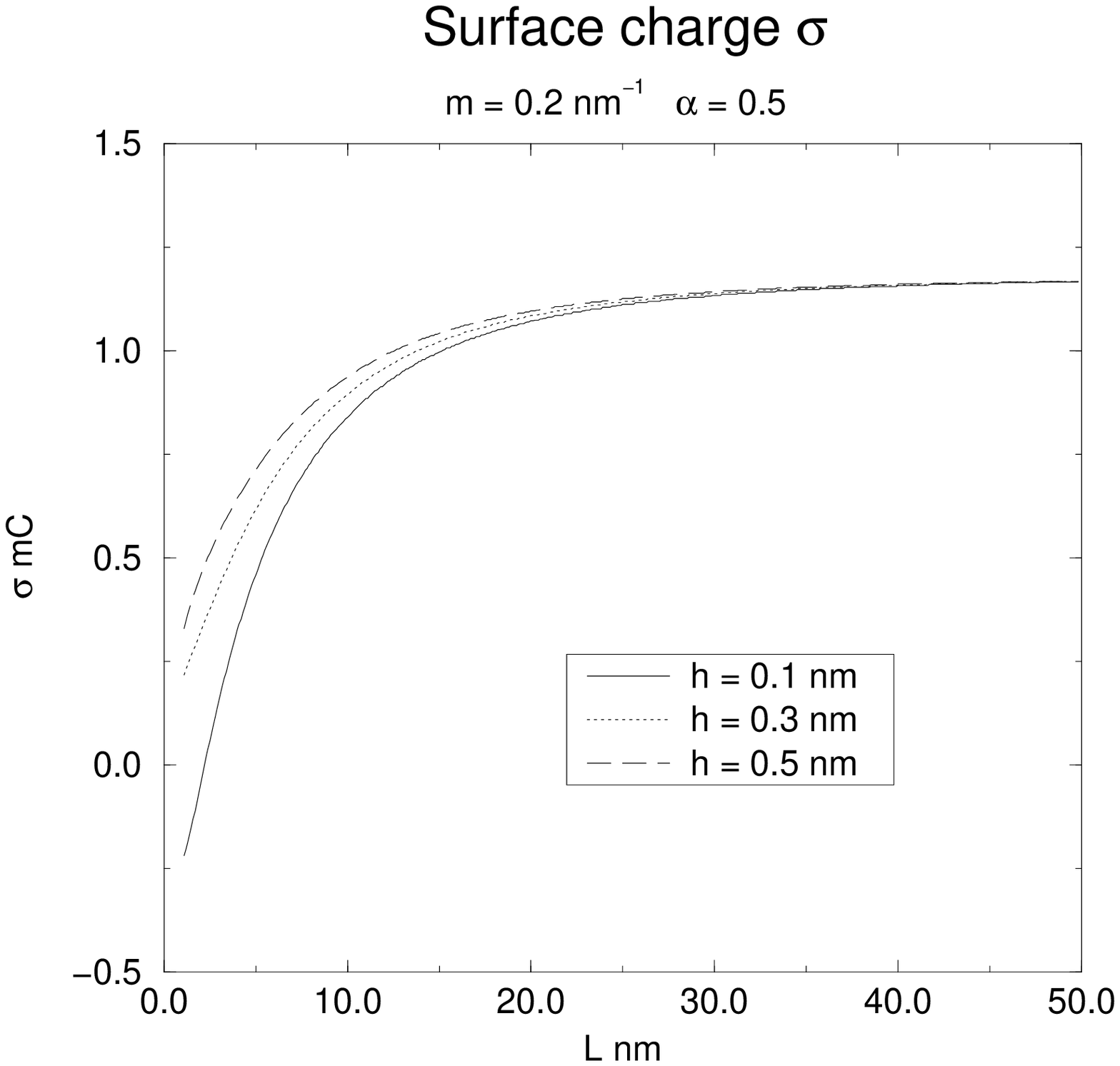,height=90mm}
\enc
\caption{\label{sig_h}\small
The surface charge $\sigma$ in millicoulombs in Eq. (\ref{sigl}) for Debye mass
$m=0.2 \mbox{nm}^{-1}$ and $\alpha = 0.5$ for different values of the thickness
of the Stern layer $h = 0.1, 0.3, 0.5~\mbox{nm}$. We see that $\sigma$ decreases with $L$
and this effect is enhanced as $h$ becomes smaller as we should expect since the image
charges have greater influence. However, even though $\sigma$ is constrained to vanish we
see that for small enough $h$ it changes sign and so must have minimum at very small $L$
before turning towards zero. This effect is evident in Fig (\ref{sig_h_mf_1loop}). 
Whether or not this effect is an artifact will be a subject of further study.
}
\enf

\befh
\bec
\epsfig{file=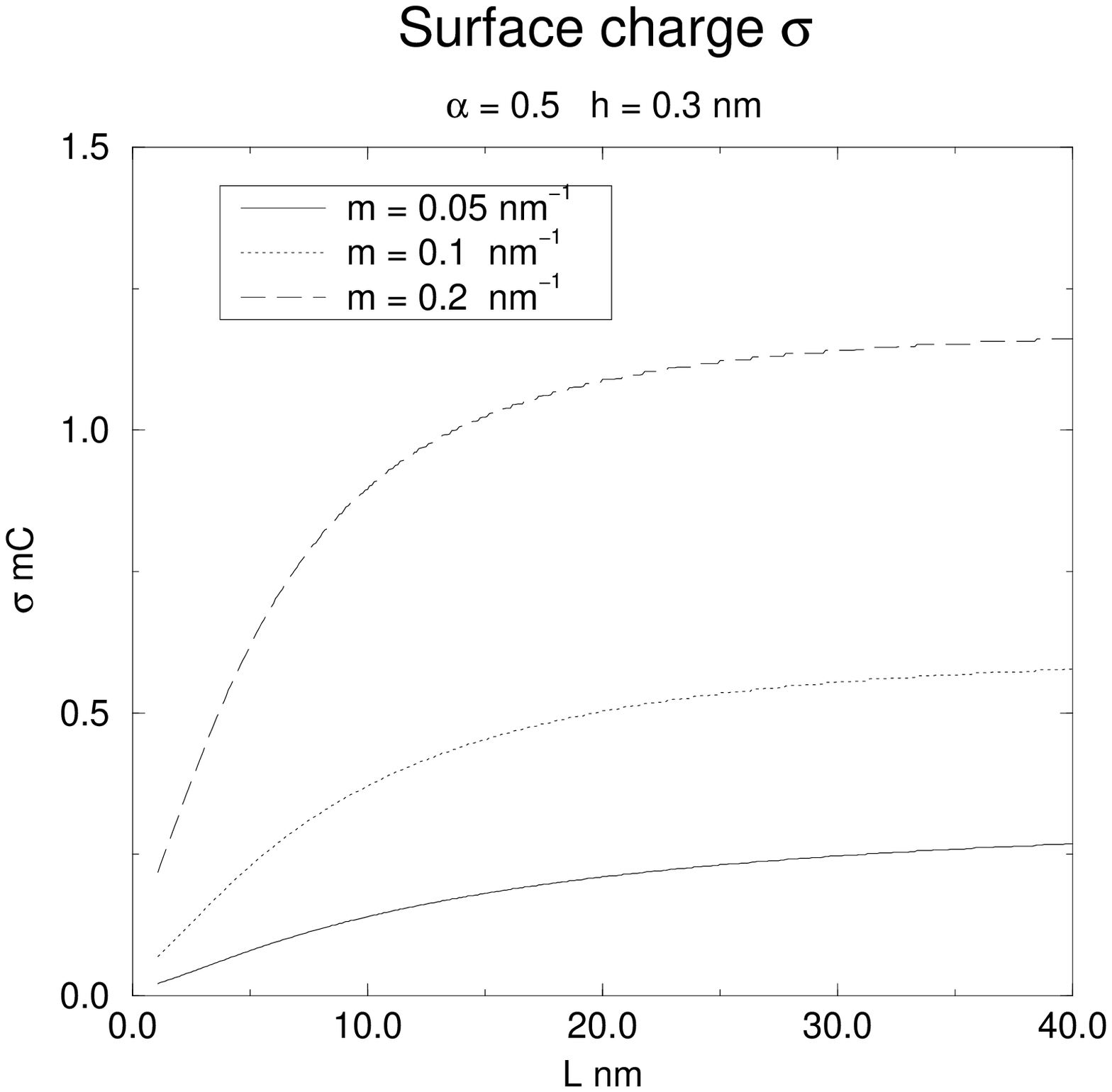,height=90mm}
\enc
\caption{\label{sig_m}\small
The surface charge $\sigma$ in millicoulombs in Eq. (\ref{sigl}) for $\alpha=0.5$ and Stern layer
thickness  $h = 0.3~\mbox{nm}$ for values of the Debye mass $m=0.05,0.1,0.2~\mbox{nm}^{-1}$.
}
\enf

\befh
\bec
\epsfig{file=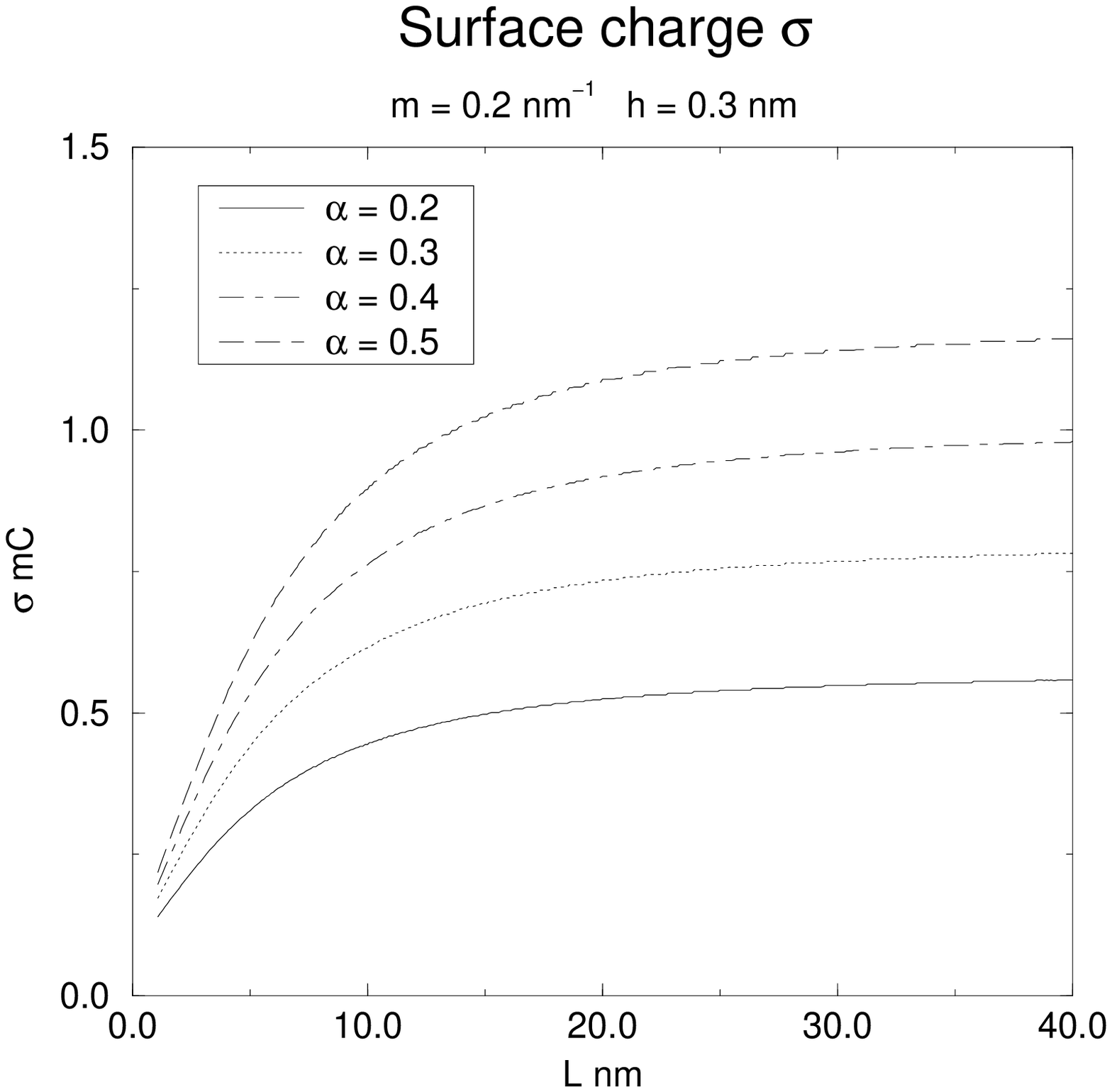,height=90mm}
\enc
\caption{\label{sig_a}\small
The surface charge $\sigma$ in millicoulombs in Eq. (\ref{sigl}) for Debye mass 
$m = 0.2~\mbox{nm}^{-1}$ and Stern layer thickness  $h = 0.3~\mbox{nm}$ for values 
of $\alpha = 0.2,0.3,0.4,0.5$.
}
\enf

\befh
\bec
\epsfig{file=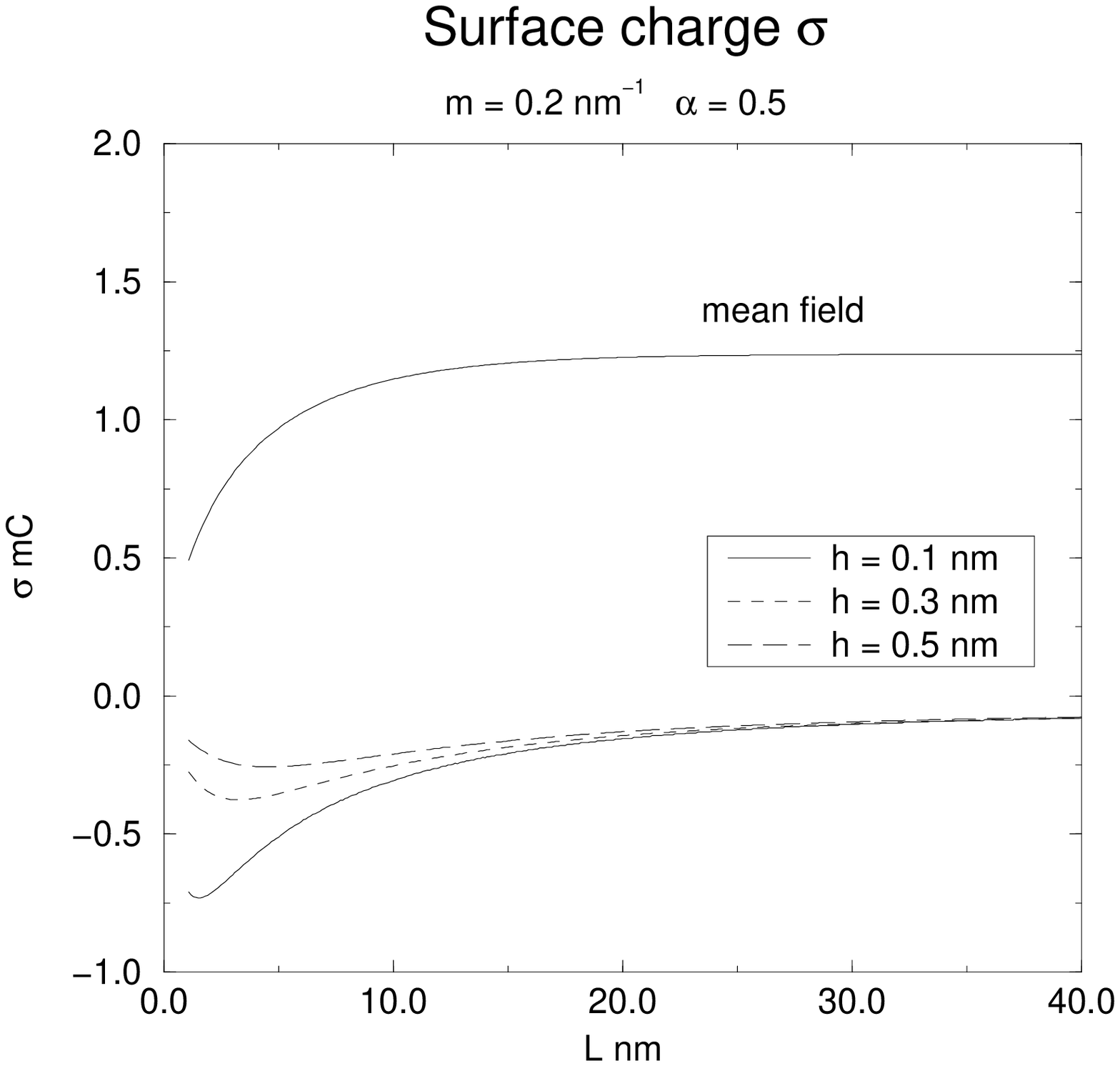,height=90mm}
\enc
\caption{\label{sig_h_mf_1loop}\small
The surface charge $\sigma$ in millicoulombs in Eq. (\ref{sigl}) showing the mean-field
contribution and the one-loop ($O(g)$) contributions as a function of Stern layer thickness
$h = 0.1, 0.3, 0.5~\mbox{nm}$. The one-loop contribution is negative 
and has a minimum before
turning to zero as it must since $\alphap(L)$ vanishes as $L \to 0$.
}
\enf

\end{document}